\documentclass[%
 aip,rsi,amsmath,amssymb,reprint,longbibliography]{revtex4-1}

\usepackage{graphicx}
\usepackage{dcolumn}
\usepackage{bm}
\usepackage{gensymb}
\usepackage{color}

\begin{document}

\preprint{AIP/123-QED}

\title{Detection of the microwave emission from a spin-torque oscillator by a spin-torque diode}

\author{Danijela Markovi\'c}
\affiliation{Unité Mixte de Physique CNRS/Thales, Universit\'e Paris-Saclay, 91767 Palaiseau, France
}

\author{Nathan Leroux}
\affiliation{Unité Mixte de Physique CNRS/Thales, Universit\'e Paris-Saclay, 91767 Palaiseau, France
}

\author{Alice Mizrahi}
\affiliation{Unité Mixte de Physique CNRS/Thales, Universit\'e Paris-Saclay, 91767 Palaiseau, France
}

\author{Juan Trastoy}
\affiliation{Unité Mixte de Physique CNRS/Thales, Universit\'e Paris-Saclay, 91767 Palaiseau, France
}

\author{Vincent Cros}
\affiliation{Unité Mixte de Physique CNRS/Thales, Universit\'e Paris-Saclay, 91767 Palaiseau, France
}

\author{Paolo Bortolotti}
\affiliation{Unité Mixte de Physique CNRS/Thales, Universit\'e Paris-Saclay, 91767 Palaiseau, France
}

\author{Leandro Martins}
\affiliation{International Iberian Nanotechnology Laboratory (INL), 4715-31 Braga, Portugal}

\author{Alex Jenkins}
\affiliation{International Iberian Nanotechnology Laboratory (INL), 4715-31 Braga, Portugal}

\author{Ricardo Ferreira}
\affiliation{International Iberian Nanotechnology Laboratory (INL), 4715-31 Braga, Portugal}

\author{Julie Grollier}
\affiliation{Unité Mixte de Physique CNRS/Thales, Universit\'e Paris-Saclay, 91767 Palaiseau, France
}

\begin{abstract}

Magnetic tunnel junctions are nanoscale spintronic devices with microwave generation and detection capabilities. Here we use the rectification effect called "spin-diode" in a magnetic tunnel junction to wirelessly detect the microwave emission of another junction in the auto-oscillatory regime. We show that the rectified spin-diode voltage measured at the receiving junction end can be reconstructed from the independently measured auto-oscillation and spin diode spectra in each junction. Finally we adapt the auto-oscillator model to the case of spin-torque oscillator and spin-torque diode and we show that accurately reproduces the experimentally observed features. These results will be useful to design circuits and chips based on spintronic nanodevices communicating through microwaves.
\end{abstract}

\maketitle

\section{Introduction}

Spin-polarized direct electric current $I_{dc}$ in magnetic tunnel junctions exerts a spin-transfer torque on the magnetization of the free layer and can induce its auto-oscillations \cite{Slonczewski1996, Berger1996}. This process is fundamental in spin-torque nano-oscillators \cite{Kiselev2003, Rippard2004} where magnetization dynamics are sustained and converted to voltage oscillations in the GHz range. Spin-torque nano-oscillators are promising as microwave emitters for future wireless applications due to their nanoscale dimensions, their large signal to noise ratio, and their ability to modulate and demodulate signals \cite{Pufall2005, Wickenden2009, Choi2014, Ruiz-Calaforra2017, Litvinenko2019}. When a magnetic tunnel junction is driven by a microwave current $I_{RF}(t) = I_{RF}\sin{(\omega t)}$, it leads to the spin-diode effect. The spin-polarized microwave current exerts an oscillating spin torque on the magnetization of the free layer, leading to resonant magnetization precessions and consequently to resistance oscillations of the junction $R(t)$. The oscillating resistance partly rectifies the current such that a dc voltage $V_{SD} = \langle I_{RF}(t) R(t) \rangle$ builds across the junction. Since it was measured for the first time \cite{Tulapurkar2005}, the spin diode effect was studied for energy harvesting \cite{Fang2019, Prokopenko2012}, time-resolved measurements of resistance oscillations \cite{Wang2011a} and microwave detection \cite{Fang2016, Miwa2013, Jenkins2016}. Yet, the fundamental building block for future all-spintronic wireless communication network, consisting of a spin oscillator emitter and spin diode detector has not been demonstrated to date. The first difficulty is to match their frequencies so that the oscillator can influence the diode behaviour, which requires to use different diameter junctions for the oscillator and the diode. The second difficulty is to analyze and understand the output signal at the detector.

In this letter we report the detection of the microwave signal emitted in free air by a spin-torque nano-oscillator with the spin-diode effect in a magnetic tunnel junction. We first measure the oscillator emission and the spin diode rectification separately. We then make the oscillator and the diode communicate through microwave antennas and study the evolution of the spin-diode signal as a function of the dc current in the oscillator as well as the amplification gain in the line. We show that we can accurately  
\begin{figure}
\includegraphics[scale=0.4]{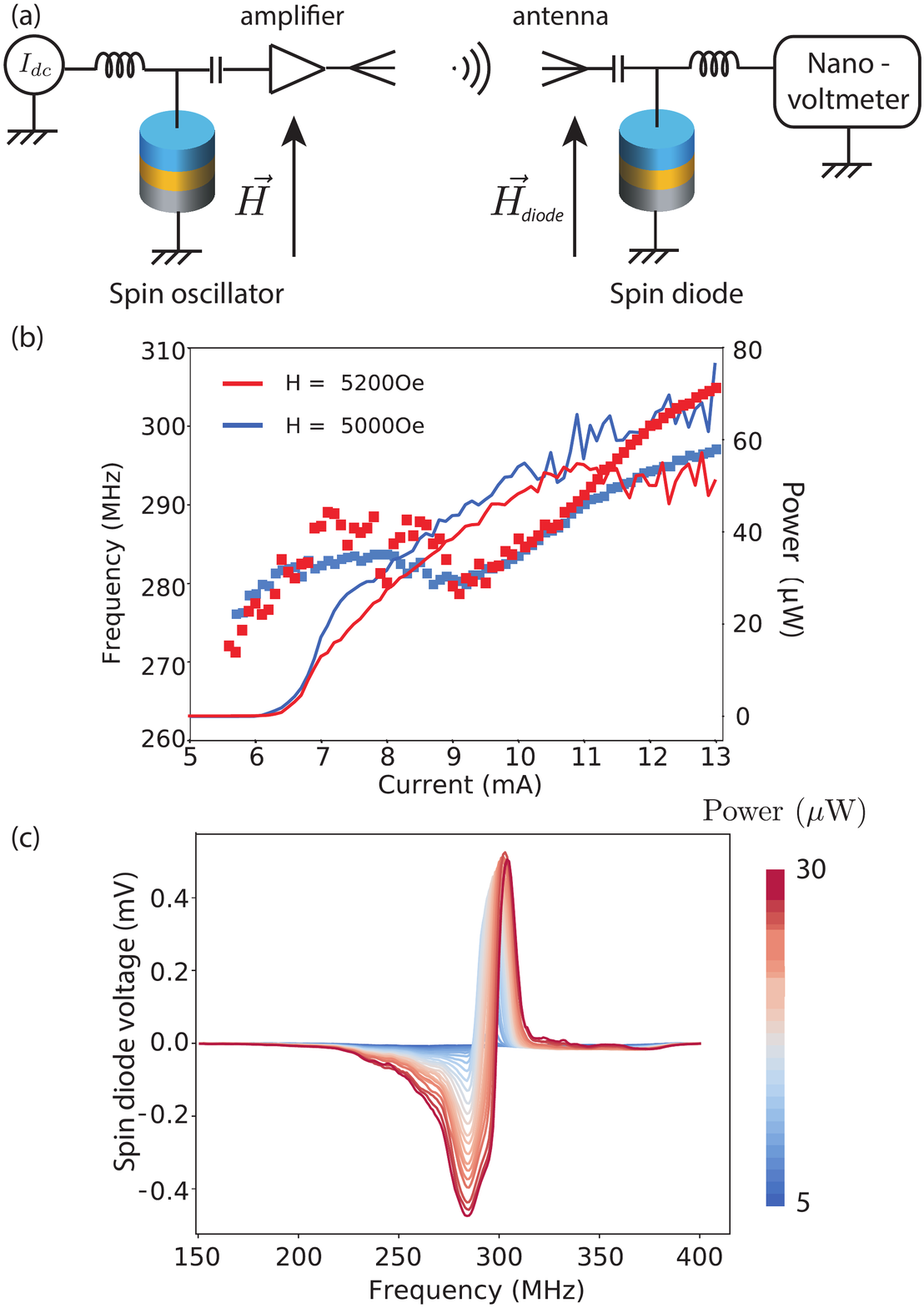}
\caption{(a) Schematic of the measurement setup. Both the spin-torque nano-oscillator and spin-torque diode are magnetic tunnel junctions, composed of two ferromagnets separated by a thin non-magnetic layer (yellow). The magnetization of the bottom ferromagnet (grey) is pinned, whereas that of the top one (blue) is free. (b) Frequency (squares) and power (line) of the oscillator after 20 dB total gain after the amplifier and antennas as a function of current $I_{dc}$ extracted from the measured power spectra for two different oscillator fields $H = 5200$ Oe (red) and $H = 5000$ Oe (blue). (c) Measured rectified spin diode voltage at $H_{diode} = 3000$ Oe as a function of the microwave source frequency for different color coded source powers. }
\label{Figure1}
\end{figure}
reconstruct and understand the measured spin-diode signal by combining the independently measured signals from the oscillator and the diode. Finally we use the auto-oscillator model \cite{Slavin2009} and we solve it analytically and numerically for the case of the vortex spin-torque oscillator and spin-torque diode. We find that it accurately accounts for the experimentally measured features.

\section{Experimental setup and device parameters}

Both the spin-torque oscillator and the spin-torque diode used in this experiment are nano-pillars composed of a 1.8 nm thick CoFeB layer whose magnetization is pinned by an underlying synthetic antiferromagnet, 1 nm thick MgO insulating barrier and 7 nm thick NiFe free magnetic layer whose ground state is a magnetic vortex. The spin-torque oscillator used as an emitter is biased with a direct current to induce auto-oscillations. In the case of vortex oscillators the oscillation frequency increases strongly with the injected direct current $I_{dc}$ compared to the base resonant frequency $\omega_{osc}^0$ due to the Oersted field \cite{dussaux_field_2012}
\begin{equation}
    \omega_{osc} = \omega_{osc}^0 + \kappa^{Oe}I_{dc} + N_{osc}p_{osc},
\end{equation}
where $\kappa^{Oe}$ is a parameter accounting for the Oersted field-induced confinement and $p_{osc}$ and $N_{osc}$ are the oscillator power and nonlinear frequency shift. The spin-torque diode used as a detector responds to RF currents within a frequency range around its resonant frequency. As there is no direct current injected in the diode, its frequency evolution is given by 
\begin{equation}
    \omega_{d} = \omega_{d}^0 + N_{d}p_{d},
\end{equation}
where $\omega_{d}^0$ is the diode resonance frequency, $p_d$ its oscillation power and $N_d$ its non-linear frequency shift. This means that the base resonant frequency of the oscillator should be lower than the resonance frequency of the spin diode for their operation frequencies to match. The resonant frequency of vortex oscillators depends inversely on the diameter of the junction. Here we use a diameter of 400 nm (R = 24 $\Omega$) for the oscillator junction and a diameter of 250 nm  (R = 50 $\Omega$) for the diode junction. We furthermore independently tune the magnetic field applied to each junction to modify their individual frequencies.

 The schematic of the experimental setup is shown in Figure \ref{Figure1}(a).  We begin by measuring the spin-torque oscillator alone. We apply a magnetic field $\vec{H}$ perpendicular to the magnetic stack to cant the magnetization of the pinned reference layer with respect to that of the free layer, which is important to create a spin torque with the proper symmetry to induce sustained vortex gyrations \cite{dussaux_field_2012}. When the injected direct current $I_{dc}$ is larger than a critical threshold value $I^{th}$, the spin torque is sufficient to induce oscillations of the vortex core. We measure the resulting microwave voltage across the junction $V(t)$ with a Power Spectrum Analyser. The frequency (squares) and the power (lines) of the oscillator after an amplification of 22 dB are plotted as a function of current $I_{dc}$ in Figure \ref{Figure1}(b) for two different applied magnetic fields $H = 5200$ Oe and $H = 5000$ Oe. The obtained trends reveal different oscillation regimes depending on the current. Above the threshold current $I^{th} = 6.38$ mA and below 9 mA, the power of the oscillator is low and we observe (not shown) that the linewidth is large, above 30 MHz, with SNR $\approx 10^4$. This mode probably corresponds to thermal excitations of a pinned vortex below the critical current threshold \cite{Dussaux2010a}. Above 9 mA, both the power and frequency depend quasi-linearly on current as expected from strong gyrotropic vortex precessions and can be tuned with the magnetic field as well. In this regime, the linewidth is reduced below 1 MHz, with SNR $\approx 10^6$.
 
\begin{figure}
\includegraphics[scale=0.35]{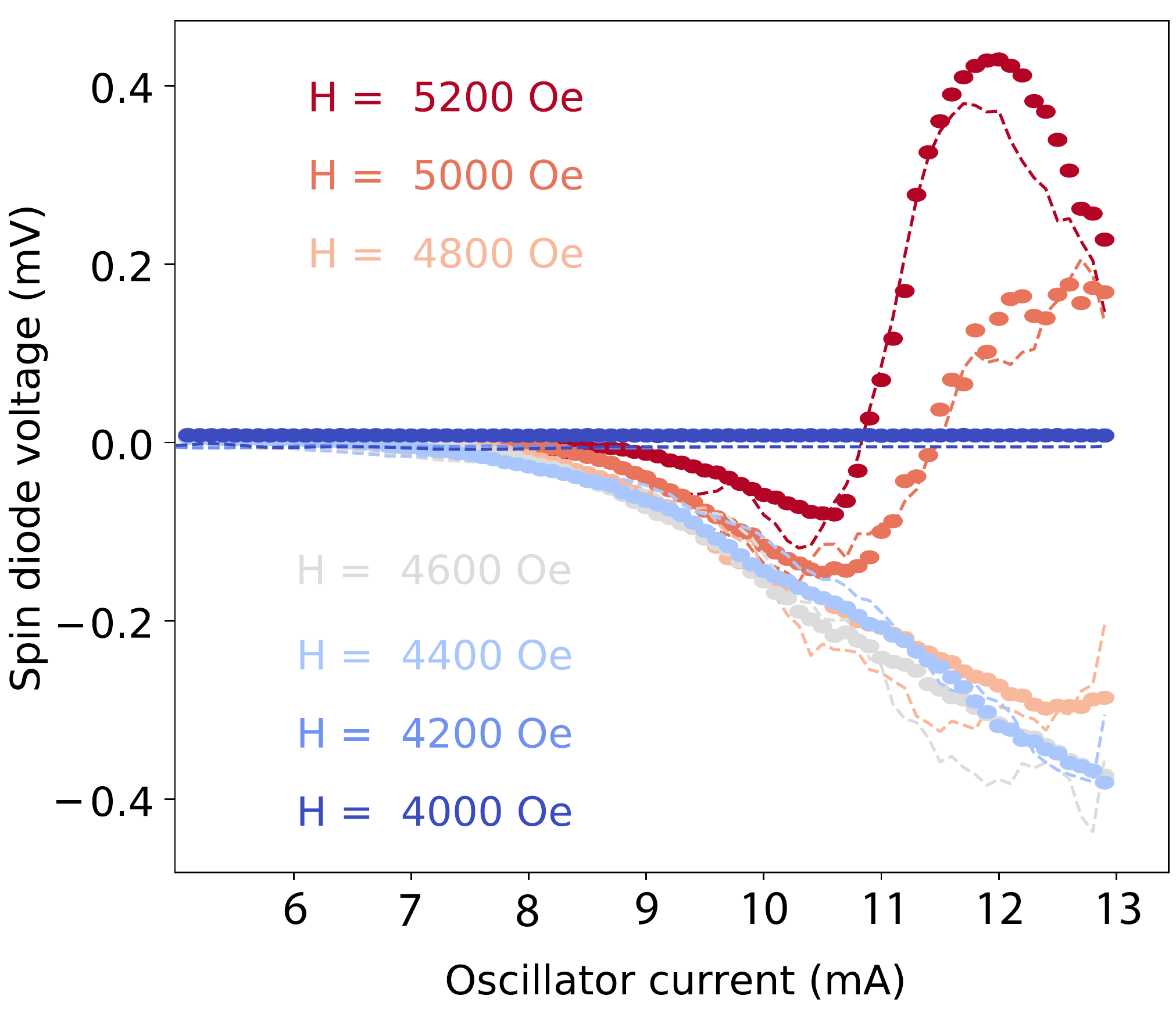}
\caption{Measured spin diode voltage as a function of the current in the oscillator for different oscillator fields $H$. Dashed lines show the voltage reconstructed from the oscillator amplitude at a given current (shown for $H$ = 5200 Oe and $H$ = 5000 Oe in Figure~\ref{Figure1}(b)) and spin diode voltage at a given frequency (Figure~\ref{Figure1}(c)).}
\label{Figure2}
\end{figure}

\section{Results}
 
We first study the spin-torque diode alone by applying to it a microwave current $I_{RF}(t)$. The experiment is performed in the presence of a perpendicular magnetic field $H_{diode} = 3000$ Oe chosen to bring the spin diode resonance within the frequency range of the oscillator. The voltage measured as a function of the microwave frequency is shown in Figure \ref{Figure1}(c) for different microwave source powers. The shape of the power spectra is mainly anti-Lorentzian, corresponding to a dominant contribution of the field-like torque terms \cite{Sankey2008, Kubota2008}. Due to the oscillator nonlinearity, the resonance frequency increases with increasing power \cite{chen_spin-torque_2009, Slavin2009}. The SNR is typically $100$ for the microwave powers that we send.

In a second step, we use the spin-torque diode effect in the second magnetic tunnel junction to detect the oscillator emission from the first junction. The amplified oscillator voltage is injected into a wireless transceiver RF antenna with mean attenuation of 7.3 dB in the diode resonance range and lower and higher cutoff frequencies of respectively 240 and 360 MHz. The signal is received by another similar antenna that is placed at 1 cm distance from the first one in order to minimize the losses and avoid the need for a second amplifier. Typically this corresponds to maximum distance for on-chip communication. The receiver antenna converts it to a microwave current that is injected into the spin-torque diode, resulting in the spin diode voltage plotted in Figure~\ref{Figure2}. We first focus on the red curve at $H$ = 5200 Oe showing the evolution of this signal as a function of the current in the oscillator. For this field, the threshold current $I^{th}$ is 9 mA. For currents below $I^{th}$, the curve is confined close to zero because the power of the oscillator is too small and its linewidth is too large to induce a large spin-diode signal. Above $I^{th}$, the power increases, the linewidth decreases and a spin-diode signal appears. The oscillator frequency increases with the input current, and spans from 280 to 305 MHz across the spin diode resonance. The resulting spin-diode voltage has an anti-Lorentzian-like shape as a function of the oscillator current similar to the measurements of Fig.~\ref{Figure1}(c). However this shape is distorted because the power emitted by the oscillator is not constant. The oscillator power increases with current, so that the absolute value of the spin-diode voltage is enhanced at large currents compared to Fig.~\ref{Figure1}(c). As can be seen from Fig.~\ref{Figure1}(b), the oscillator frequency range that is spanned with current depends on the magnetic field $\vec{H}$. The maximum frequency of the oscillator decreases with the magnetic field that we vary from 4000 to 5200 Oe. For lower fields, higher currents are therefore needed to reach the spin-diode resonance and only part of the curve can be spanned  before reaching currents (above 15 mA) that damage the device.

\begin{figure}
\includegraphics[scale=0.35]{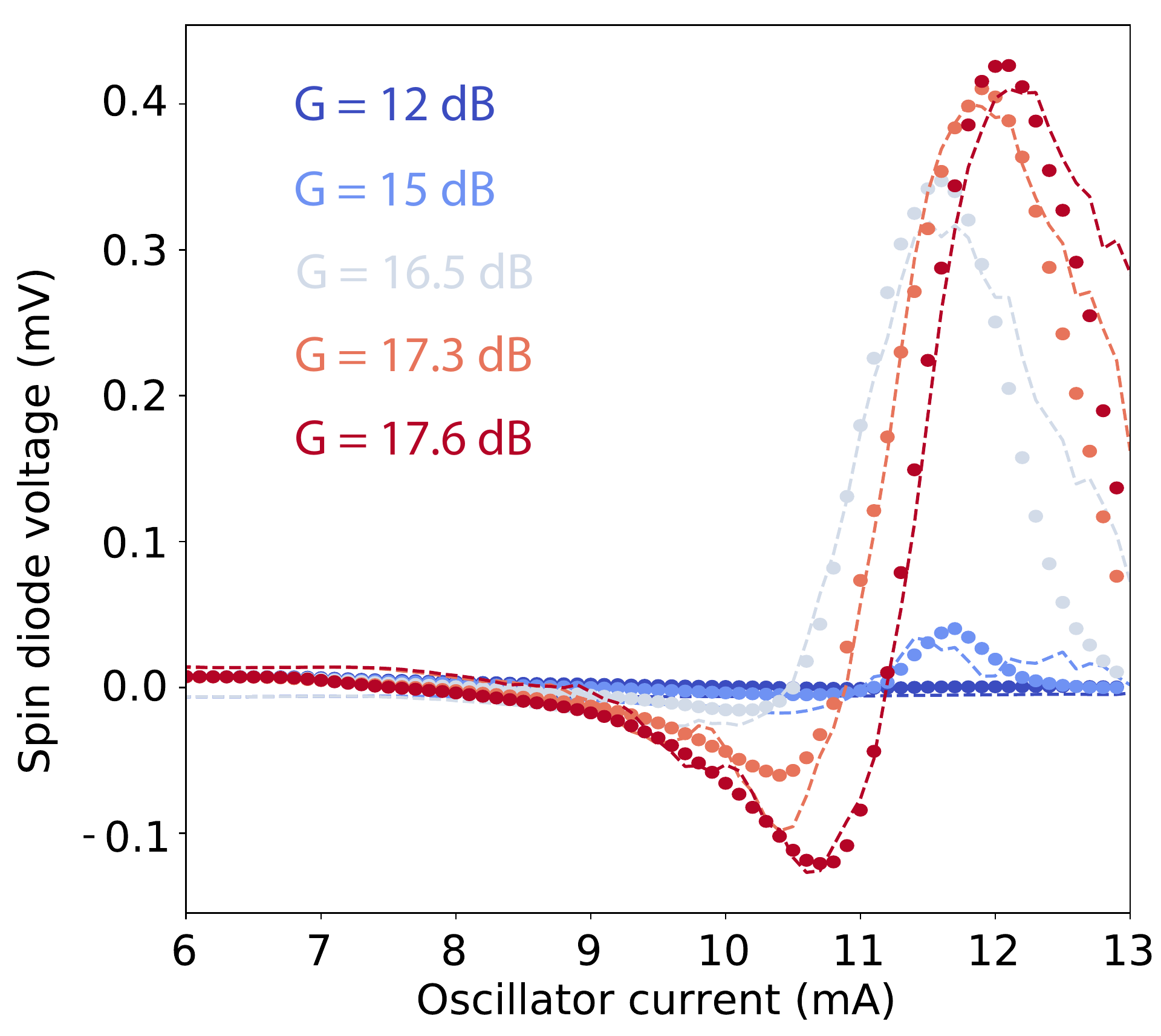}
\caption{Measured spin-diode voltage as a function of the oscillator current for fixed oscillator magnetic field $H$ = 5200 Oe and for different amplifier gains G. Dashed lines correspond to voltages reconstructed from the oscillator amplitude at a given current (Figure~\ref{Figure1}(b)) and spin diode voltage for given input frequency and power (Figure~\ref{Figure1}(c)).}
\label{Figure3}
\end{figure}

We then study the impact of amplification gain on the detected signal. Figure \ref{Figure3} shows the measured spin-diode voltage for different amplifier gains at a fixed oscillator field of $H$ = 5200 Oe. As expected, the detected signal amplitude increases with the gain. Furthermore, the curve shifts with amplification, which comes from the diode non-linearity already observed from the resonance frequency shift in Figure~\ref{Figure1}(c). Such gain-dependent shift will need to be taken into account in circuits leveraging wireless transmission between spintronic oscillators and diodes.

Towards this goal, we first check if we can reconstruct the signals in Figs.~\ref{Figure2} and \ref{Figure3} detected at the spin-diode end from the individual oscillator and diode characterizations. For a given current in the oscillator, we determine the oscillator frequency and emitted power from the measured emission spectra (Figure~\ref{Figure1}(b)). The input power in the diode is equal to the power emitted by the oscillator, amplified by a value that depends on the chosen amplifier gain, as well as the losses in the antennas and the RF cables at a given frequency. The total gain of the chain composed of the amplifier, microwave cables and antennas is a priori known but it is not flat in frequency. The losses induced by the cables and microwave components are (3$\pm$0.5) dB and by the antennas (7.3$\pm$0.5) dB. We thus adjust its value by matching the reconstructed output to the experimental data. As shown in Figure~\ref{Figure2} and Figure~\ref{Figure3}, this reconstructed voltage, in dashed line, matches very well the wirelessly transmitted signal. 

\begin{figure}[!h]
\includegraphics[scale=0.5]{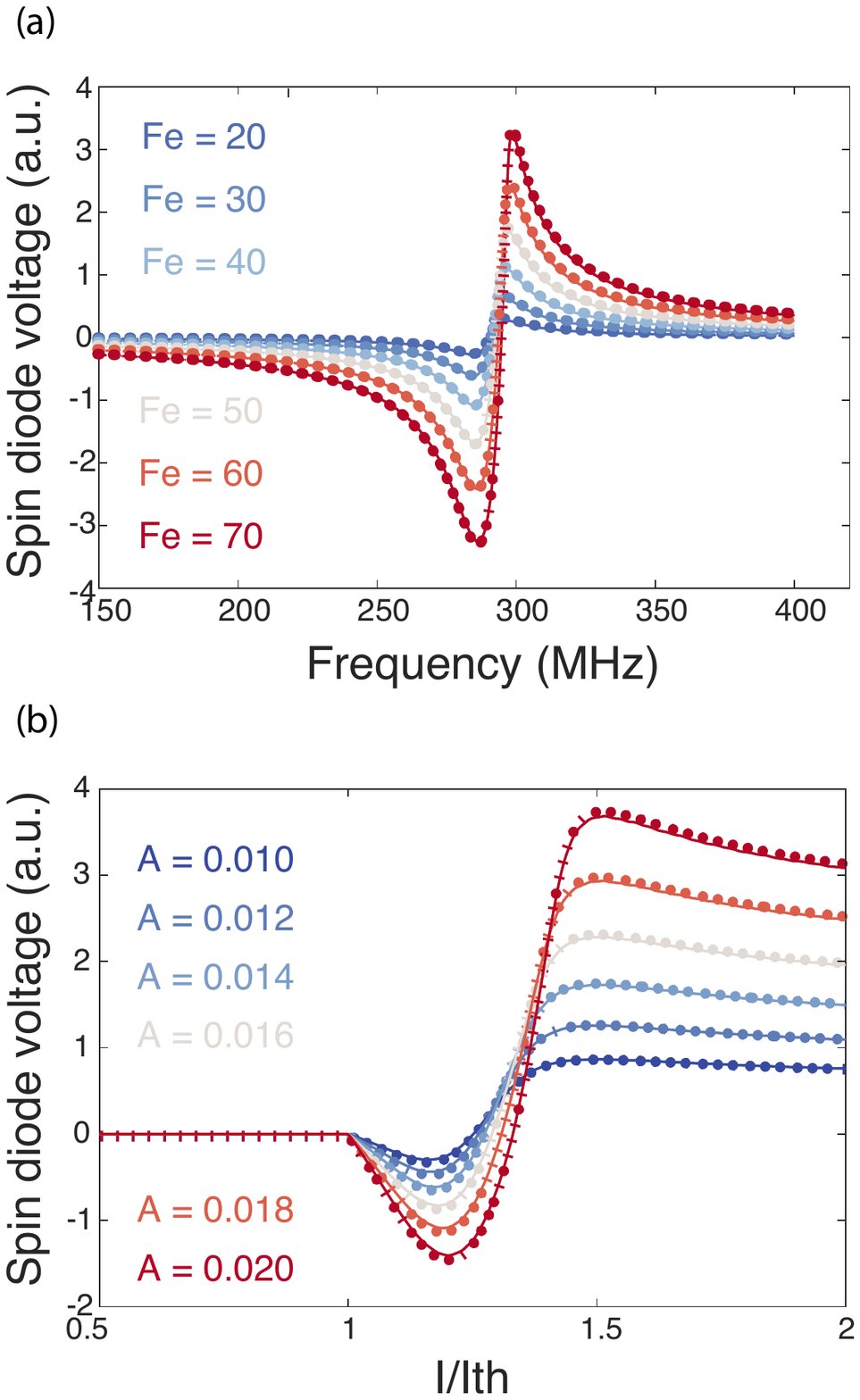}
\caption{(a) Calculated spin diode voltage as a function of source frequency for different source powers and (b) as a function of oscillator current for different amplifying factors obtained by numerically solving dynamics equation Eq.~\eqref{eq_dynamique} (full line) and from the exact analytical solution Eq.~\eqref{eq_analytic} (dashed line).}
\label{Figure4}
\end{figure}

\section{Analytical model and numerical simulations}

We now show that this signal can be analytically modelled, which will be useful to design of telecommunication circuits using these elements. We describe magnetization dynamics in the oscillator and diode junctions with the non-linear dynamical equation \cite{Slavin2009}:
\begin{equation}
    \frac{dc}{dt} = (-i\omega + \Gamma_+) |c|^2 c + F_e e^{-i(\omega_e t)}, 
    \label{eq_dynamique}
\end{equation}
where $c = \sqrt{p} e^{i\phi}$ is the complex amplitude of the oscillations in the considered junction, $p$ is the normalized oscillation power, $\phi$ is its phase and $\Gamma_+$ is the net damping rate. $F_e$ accounts for the microwave force exerted on the diode. It is set to zero in the oscillator junction. We assume that the damping rate depends on power only to the first order:
\begin{equation}
    \Gamma_{+}(p) = \Gamma_{G}(1+ Q p),
    \label{eq_damping}
\end{equation}
where $\Gamma_G = \omega_0 \alpha_G$, $\alpha_G$ is the Gilbert damping and $Q$ accounts for non-linearity in damping and magnetization confinement.

We first derive the spin-diode signal from these equations. We find that for a vortex oscillator, the spin diode voltage is given by (see Appendix I for details):
\begin{equation}
    V_{SD} \propto p_d(\delta - \omega_{d}^0N_d p_d),
    \label{VSD}
\end{equation}
where $\delta = \omega_{e}-\omega_d^{0}$ is the frequency difference between the source and the diode.

For the spin-diode junction, the exact stationary solution of Eq.~\eqref{eq_dynamique} for the oscillation power is given by (see Appendix I for details)
\begin{equation}
    (Q_d^2\Gamma_{G}^2+(\omega_d^0)^2N_d^2)p_d^3+2(Q_d\Gamma_{G}^2-\omega_d^0N_d\delta)p_d^2 + (\Gamma_{G}^2 + \delta^2)p_d = F_{e}^2.
\label{eq_analytic}
\end{equation}

\begin{table}[ht]
\centering
\caption{Table of parameters used in the simulations. The
index $osc$ refers to the oscillator, and $d$ to the diode.}
\begin{center}
\begin{tabular}{ |c|c|} 
 \hline
 $\alpha_d=\alpha_{osc} = \alpha_G$ & 0.02  \\
 \hline 
 $N_d=N_{osc}$ & 0.5  \\ 
  \hline 
 $Q_d=Q_{osc}$ & 2  \\ 
   \hline 
 $\omega_d^0$ & 2 $\pi \times$ 290 MHz  \\ 
    \hline 
$\omega_{osc}^0$ & 2 $\pi \times$ 280 MHz  \\ 
     \hline 
\end{tabular}
\end{center}
\label{table}
\end{table}%

We plot in Figure~\ref{Figure4}(a) in dashed lines the spin diode voltage $V_{SD}$ calculated from Eq.~\eqref{eq_analytic} as a function of microwave source frequency for different external drive forces $F_e$. We superimpose in full lines the numerical solution directly extracted from Eq.~\eqref{eq_dynamique} using a fourth order Runge-Kutta method. The parameters that we used are given in Table~\ref{table}. Note that $c$ being a dimensionless amplitude, the driving force $F_e$ is homogeneous to a frequency. Figure~\ref{Figure4}(a) shows that the agreement between numerical and analytical solutions for the spin-diode signal is excellent. Furthermore, the experimental trends in Figure~\ref{Figure1}(c) are well reproduced: the shape of the spin-diode signal is anti-Lorentzian, its amplitude increases and the resonant frequency shifts with increasing power. 

Next, we simulate the spin diode voltage as a function of the oscillator current for different amplification factors in Figure~\ref{Figure4}(b). The oscillator Gilbert damping $\alpha_{osc}$ and nonlinear coefficients for frequency $N_{osc}$ and damping $Q_{osc}$ are the same as the spin diode. The normalized oscillator power can be calculated from Eq.~\eqref{eq_dynamique} at steady state with $F_e = 0$,
\begin{equation}
\begin{tabular}{ c c c} 
${p}_{osc} = \frac{\xi-1}{\xi+Q_{osc}}$  \ if \ $\xi > 1$\\
${p}_{osc} = 0$ \ if \ $\xi \leq 1$,
\end{tabular}
\end{equation}
where $\xi = \frac{I}{I^{th}}$. We consider that the driving force $F_e$ on the diode is proportional to the amplitude of the spin-torque oscillator 
\begin{equation}
    F_{e} = A\sqrt{p_{osc}},
\end{equation}
where $A$ is a factor that depends on the overall amplification gain in the line and the spin-transfer torque efficiency.

We observe in Figure~\ref{Figure4}(b) a behavior that is qualitatively similar to the measurements. As expected, we observe the frequency shift due to the diode nonlinearity and diode voltage of larger amplitude for larger currents due to the oscillator power dependence on current. Our calculation does not take into account the bias-dependence of the Tunnel Magneto-Resistance in the oscillator, which explains why the spin-diode signal is larger at high currents in the predictions compared to the experiments. In the future, quantitative agreements can be obtained if necessary by refining the model to take into account experimental details such as amplification losses in the cables and line, bias dependence of Tunnel Magneto-Resistance and spin-transfer torque, as well as accounting for torques with different symmetries and strenghs in   Eq.~\eqref{eq_dynamique}. 

In conclusion, we have wirelessly connected a spin-torque nano-oscillator to a spin-torque diode, and demonstrated that the spin diode can detect the spin oscillator microwave emission. We have reconstructed the measured rectified voltage from the independently measured auto-oscillation and spin diode spectra and shown that we can predict well the measurement result. We have adapted the auto-oscillator model for the case of vortex spin-torque oscillator and spin-torque diode and shown that it accounts for experimental trends. Our work paves the road towards realization of all-spintronic communication networks wirelessly connecting magnetic tunnel junction-based RF oscillators and diodes. Furthermore, as the sign of the spin diode voltage depends on the input frequency that can be lower or higher than spin diode resonance, a chain of such spin diodes with different resonance frequencies could be used to determine the input frequency. Performance of such oscillator frequency detector could be improved by increasing the diodes signal to noise ratio and TMR, and reducing their bandwidth. 

This work was supported by the European Research Council ERC under Grant bioSPINspired 682955, the French ANR project SPIN-IA (ANR-18-ASTR-0015) and the French Ministry of Defense (DGA). The authors acknowledge Ursula Ebels and Artem Litvinenko for fruitful discussions and help with RF transducer antennas.

\section{Appendix I}
The universal model for magnetization dynamics under the influence of DC and AC signals is given by Eq.~\eqref{eq_dynamique}. This complex equation can be rewritten as two real equations describing the power $p = |c|^2$ and the phase $\phi = \arg(c)$ of the junction \cite{Slavin2009}
\begin{equation}
    \frac{dp}{dt} = -2\Gamma_{+}(p)p+2\sqrt{p}F_{e}\cos(\phi +\omega_{e}t-\psi_{e})
    \label{eq:power}
\end{equation}
\begin{equation}
    \frac{d\phi}{dt} = -\omega(p)-\frac{F_{e}}{\sqrt{p}}\sin(\phi +\omega_{e}t-\psi_{e}).
    \label{eq:phase}
\end{equation}
We concentrate here on the case of the diode, which only receives an AC input. We introduce the phase difference between the harmonic signal and the diode $\Phi = \phi +\omega_{e}t-\psi_{e}$, such that using Eq.~\eqref{eq:phase} we find
\begin{equation}
    \frac{d\Phi}{dt} =\delta- \omega_{d}^0N_dp_d-\frac{F_{e}}{\sqrt{p_d}}\sin(\Phi),
    \label{eq:phaseDiff2}
\end{equation}
where $\delta = \omega_{e}-\omega_{d}^0$ is the frequency difference between the oscillator and the diode resonance. We measure the spin-diode voltage over few ms, which is much longer than both the oscillation period (several ns) and the relaxation time (few $\mu$s). We can thus consider to be in the steady state where the phase difference between the diode and the harmonic signal is constant over time $\frac{d\Phi}{dt} = 0$, leading to 
\begin{equation}
    \Phi = \arcsin{\frac{\sqrt{p_d}(\delta-\omega_{d}^0N_dp_d)}{F_{e}}}.
    \label{eq:phaseDiff1}
\end{equation}
Furthermore, in the stationary regime $\frac{dp_d}{dt} = 0$, such that Eq.~\eqref{eq:power} yields $\cos{\Phi}= \frac{\sqrt{p_d}\Gamma_{+}(p_d)}{F_{e}}$ and using Eq.~\eqref{eq:phaseDiff1} and $\cos(\arcsin{x}) = \sqrt{1-x^2}$, we obtain the exact diode oscillation power Eq.~\eqref{eq_analytic}. The spin diode voltage is given by
\begin{equation}
V_{SD} = \langle I_{RF}(t)R(t) \rangle = \\ I_{RF}cos(-\omega_{e}t+\psi_e) R_{P\rightarrow AP}\sqrt{p_d}\cos(\theta),
\end{equation}
where $R_{P\rightarrow AP}$ is the resistance difference between the fully parallel state and the fully anti-parallel state, $\sqrt{p_d}$ is the amplitude of oscillations in the diode and $\theta$ is the magnetoresitance oscillation phase. The force $F_e$ is proportional to the microwave amplitude $I_{RF}$.  For a vortex magnetic tunnel junction, the phase of magnetoresistance oscillation is in phase quadrature with the phase of the vortex oscillation $\phi = \theta + C\frac{\pi}{2}$, where $C$ is the chirality of the vortex, such that
\begin{align}
    V_{SD} & \propto \langle F_{e}\sqrt{p_d}C\cos(-\omega_{e}t+\psi_e)\sin(\phi) \rangle \\ & =  F_{e}\sqrt{p_d}C(\langle\sin(\phi-\omega_{e}t+\psi_e)\rangle + \langle \sin(\Phi) \rangle)/2.
\end{align}
The first term averages out over time leading to
\begin{equation}
    V_{SD} \propto CF_e \sqrt{p_d} \sin\Phi.
    \label{Vsd}
\end{equation}
Finally using Eq.~\eqref{eq:phaseDiff1} et Eq.~\eqref{Vsd} we find Eq.~\eqref{VSD}.


%

\end{document}